\documentclass[epj]{svjour}
\usepackage{graphics}
\begin{document}
\title{Improving Multiparton Monte Carlo Tools in Hadronic Collisions\thanks{Based 
on work done in collaboration with M.L.~Mangano, M.~Moretti, R.~Pittau and A.D.~Polosa.}}
\author{Fulvio Piccinini\inst{1} 
}                     
%
%
\institute{INFN - Sezione di Pavia and  Dipartimento di Fisica Nucleare 
   e Teorica, via Bassi 6, 27100 Pavia, Italy}
\date{preprint numbers: {\tt CERN-TH/2003-267}, {\tt FNT/T-2003/12}}
\abstract{Recent work on leading order multiparton calculations for hadronic 
collisions is reviewed, 
with special emphasis on the {\tt ALPGEN} event generator. Some problems 
connected with the interface of the partonic events generated via matrix elements 
with the showering are addressed.
} 
\maketitle
\section{Introduction}
\label{intro}
In high energy hadronic collisions multijet final states 
characterize a large class of important phenomena both within and 
beyond the SM. These multiparticle final states can originate from 
hard QCD radiation processes, decay of SM massive particles 
($W$, $Z$, top quark, Higgs boson), decay of heavy supersymmetric 
or more exotic particles. In general there are two 
different approaches to simulate multijet final states: the first one 
consists in generating the simplest possible final state by means of matrix 
elements and producing additional jets by parton showering 
({\tt HERWIG}~\cite{herwig}, 
{\tt ISAJET}~\cite{isajet}, 
{\tt PYTHIA}~\cite{pythia}). 
This procedure 
works well in the soft/collinear regions but fails to describe configurations 
with several widely separated jets. 
A complementary strategy is to generate high-multiplicity 
partonic final states by means of exact matrix elements and eventually 
pass the generated events to further showering. Even if it is a 
leading-order (LO) approach, this procedure can become very difficult because of 
the complexity of the matrix element calculations with many external legs 
and of the phase-space integration. Recently there has been extensive 
activity in developing several parton-level Monte Carlo (MC) event generators, 
such as 
{\tt ACERMC}~\cite{acermc}, 
{\tt ALPGEN}~\cite{alpgen}, 
{\tt AMEGIC++}~\cite{amegic}, 
{\tt CompHEP}~\cite{comphep}, 
{\tt GRACE}~\cite{grace}, 
\\
{\tt HELAC/PHEGAS/JETI}~\cite{helac}, 
{\tt MADEVENT}~\cite{madevent}. 
In this contribution the state of the art of the {\tt ALPGEN} generator 
is reviewed, paying attention to the latest improvements. The general 
problem of interfacing a LO partonic event generator with the 
parton shower is addressed, reviewing some recent work on the subject. 

\section{The {\tt ALPGEN} event generator}
\label{alp}
The {\tt ALPGEN} library is a collection of MC codes dedicated to many 
processes relevant to high energy hadron--hadron collisions. 
The calculations are based on partonic events generated by means of exact 
LO matrix elements, obtained with the {\tt ALPHA} algorithm~\cite{alpha} 
for assigned kinematics, spin, flavour and colour configurations. The generated 
unweighted events, stored according to the Les Houches Accord \#1~\cite{LesHouches} 
format, are ready for the {\tt HERWIG/PYTHIA} evolution from partons to hadrons.
Up to now the available final states in the {\tt ALPGEN} package are: 
\begin{itemize}
\item $W + N$~jets, $Z/\gamma^* + N$~jets, $N \leq 6$,
\item $W Q \bar Q + N$~jets, $Z/\gamma^* Q \bar Q + N$~jets $(Q=c,b,t)$, $N \leq 4$,
\item $W + c + N$~jets, $N \leq 5$,
\item $n\, \, W + m\, \, Z + p\, \, {\rm Higgs} + N$~jets, 
$n + m + p \leq 8$, $N \leq 3$,
\item $m \, \, \gamma + N$~jets, $N + m \leq 8$ and $N \leq 6$, 
\item $Q \bar Q + N$~jets, $(Q=c,b,t)$, $N \leq 6$,
\item $Q \bar Q Q' {\bar Q}' + N$~jets, $(Q,Q'=b,t)$ , $N \leq 4$,
\item $N$~jets, $N \leq 6$,
\item $Q \bar Q H + N$~jets, $(Q = b, t)$, $N \leq 4$.
\end{itemize}
The limitations in the maximum number of jets is only due to the setting 
of internal parameters in the {\tt ALPHA} code, which could be changed 
to accomodate a larger number of final-state particles. 
While in the first version of {\tt ALPGEN} the top quarks 
were considered as real particles, they are now ({\tt v1.3},
in the $Q \bar Q$ and $Q \bar Q H$ processes) allowed to 
decay in the three final state fermions 
($t \to W b \to b f {\bar{f}}'$) with exact matrix element, 
thus retaining all the spin correlations 
among the top decay products. The decay is calculated in the approximation 
of on-shell top quark and $W$ boson, in order to avoid the inclusion of 
non-resonant diagrams while preserving the gauge invariance of the calculations. 
The same strategy has been implemented for the decay of the gauge boson 
in the {\tt vbjets} code, the {\tt ALPGEN} directory dedicated to 
multiboson plus jets production, 
where the matrix element for the vector boson decay into a fermionic pair 
has been introduced in the zero width approximation. 
The generation of multiboson final states requires a careful treatment of the 
widths in the propagators of the unstable particles, because they generally 
break gauge invariance, giving rise to a bad high energy behaviour of the 
cross sections~\cite{gi}. The strategy 
adopted in {\tt ALPGEN} is to calculate the matrix element with the bosonic 
widths set to zero, 
removing the events containing a vector boson with a propagator 
mass $M_0$ such that 
$\vert M_0^2 - M^2 \vert < s_0$, with 
\begin{eqnarray}
\int_{-\infty}^{M^2-s_0} ds \frac{1}{(s-M^2)^2} = 
\int_{-\infty}^{M^2} ds \frac{1}{(s-M^2)^2+\Gamma^2 M^2} \nonumber \; 
\end{eqnarray}
and 
\begin{eqnarray}
\vert M_0 - M \vert \leq \frac{\Gamma}{\pi}. \nonumber
\end{eqnarray}
For all other processes that do not involve the presence of several gauge bosons, 
the fixed-widths prescription is used. The Higgs bosons are treated as stable 
particles. Their decay to fermion pairs or to four fermions will be implemented 
in a future release of the programme. 

\section{Matching partonic event generators to parton shower}
\label{match}
\begin{figure}
\resizebox{0.45\textwidth}{!}{%
  \includegraphics{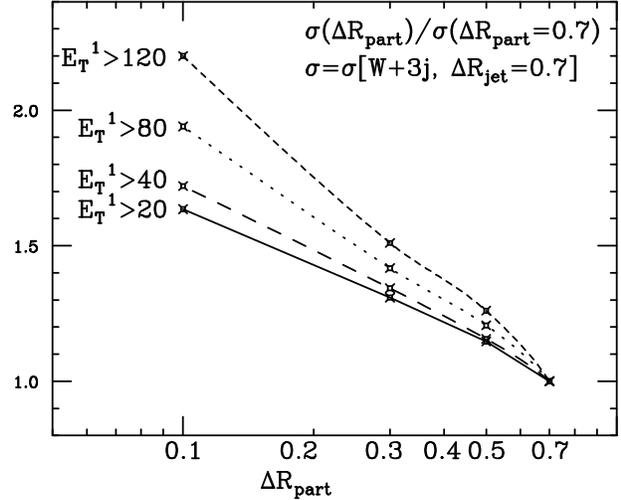}
}
\caption{The rate for $p \bar p \to W + 3$~jets at Tevatron as a function 
of the partonic separation cut $\Delta R_{\rm part}$ normalized to the 
cross section for $\Delta R_{\rm part} = 0.7$.}
\label{fig:w3j}    
\end{figure}

\begin{figure}
\resizebox{0.45\textwidth}{!}{%
  \includegraphics{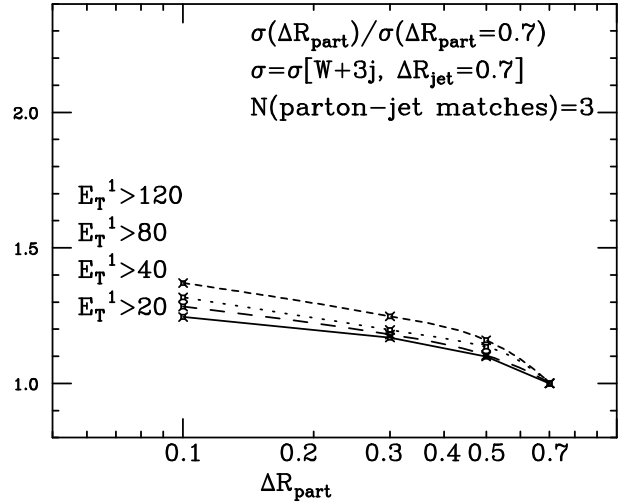}
}
\caption{The same as in fig.~\ref{fig:w3j}, requiring a jet matching for 
every parton~\cite{mlmtev}.}
\label{fig:w3jm}     
\end{figure}

In order to simulate the real hadronic final states, the partonic events have 
to be passed through a MC parton shower. 
However, in this interface there 
is some ambiguity in the cuts implemented at the partonic level. They 
are unphysical, in the sense that the final jet cross-section should be 
independent of their choice, provided that they are not harder than the 
cuts applied to the real jets. However, starting with looser partonic cuts 
increases the probability of obtaining $n$ jets from $n+m$ partons after 
parton showering (giving rise to the double-counting problem), 
as can be seen in fig.~\ref{fig:w3j}, which shows the jet rates with the 
constraint $E_T^{jet} > E_T^{cut}$ for the hardest jet in $W + 3$~jets 
events at Tevatron, versus the parton separation $\Delta R_{\rm part}$ 
imposed at the level of ME generation. The jets are reconstructed with 
the cone algorithm and the cross sections are normalized to the 
result obtained with $\Delta R_{\rm part} = 0.7$. 
The increasing ratio for smaller $\Delta R_{\rm part}$ is due to 
both the collinear divergence of 
the matrix element for $\Delta R_{\rm part} \to 0$ and the 
increasing double counting for smaller $\Delta R_{\rm part}$. 
A first approximation to the solution of the problem could consist in 
requiring a jet matching for every parton~\cite{mlmtev}. 
With this recipe the shapes of the $W + 3$~jets example above 
become flatter, as diplayed in fig.~\ref{fig:w3jm}, but still showing 
a residual dependence on the parameter $\Delta R_{\rm part}$. 
The general problem of matching multiparton matrix element calculations 
with parton showers has been extensively studied in the literature 
and for $e^+ e^-$ collisions a solution (CKKW) has been proposed and tested 
on LEP data~\cite{ckkw}, which avoids double counting and shifts the 
dependence on the resolution parameter beyond next-to-leading logarithmic (NLL) 
accuracy. The method 
consists in separating arbitrarily the phase-space regions covered 
by matrix element and parton shower, and use, for all parton multiplicities, 
vetoed parton showers together with reweighted tree-level matrix elements 
by means of suitable Sudakov form-factor combinations. 
The necessary steps for the implementation of the 
procedure can be summarized as follows: 
\begin{itemize}
\item select the jet multiplicity $n$ according to the jet rates obtained 
      with matrix elements with resolution $y_{ij} > y_{cut}$, defined 
      according to the $k_T$-algorithm~\cite{durham} 
      ($y_{ij} = 2 {\rm min}(E_i^2, E_j^2) / {\hat{s}} (1 -\cos \vartheta_{ij})$);

\item generate $n$ parton momenta according to the matrix element with 
      fixed $\alpha_s(y_{cut})$ and 
      reweight the event with the probability of no further branching by means 
      of Sudakov form factors;

\item build a ``PS history'' by clustering the partons to determine 
      the values at which 1,2,...,$n$ jets 
      are resolved. In so doing a tree of branchings is constructed and 
      the nodal scales characteristic of each branching 
      are used to reweight the event with running $\alpha_s$;

\item apply a coupling constant reweighting factor 
      $\alpha_s(y_i)$ / $\alpha_s(y_{cut}) \leq 1$ for every branching 
      of the ``PS history'', where $y_i$ is the nodal scale; 

\item after successful unweighting, use the $n$-parton kinematics as initial 
      condition for the shower, vetoing all branchings such that 
      $y_{ij} > y_{cut}$.
\end{itemize} 
The extension of the procedure to hadronic collisions has been proposed 
in ref.~\cite{krauss}. 
Recent detailed results of the implementation of the 
procedure with the programs {\tt HERWIG}, {\tt PYTHIA} and {\tt SHERPA} 
have been presented in ref.~\cite{ckkwh}. 
The implementation in {\tt ALPGEN} is currently under 
investigation. Preliminary studies regarding the partonic steps with 
{\tt ALPGEN} have been presented in ref.~\cite{mlmmcw}. 
In the case of hadronic collisions there is a certain 
degree of arbitrariness, such as the choice of the 
Sudakov form factors, the choice of the scale of $\alpha_s$ (LO or NLO), 
the treatment of the highest-multiplicity matrix element, 
the choice of the clustering scheme, the use of flavour or colour 
information to define the tree and the related reweighting factors, 
the treatment of gauge bosons. 
All these uncertainties entail that a large degree of tuning on the data 
(possibly process-dependent) will be needed, and further work remains to be done 
to find what the correct prescriptions are. 
The $\alpha_s$ reweighting of the partonic events could be important on its own, 
because it should effectively give, in a gauge-invariant way, the bulk of the NLO 
QCD corrections. This could be tested in cases where multijet NLO calculations 
are available. 

\section{Summary}
The MC simulation of hard multiparticle final states at hadronic 
colliders is a very important issue. Thanks to recent efforts 
by different groups, several  multiparton event generators based on exact 
matrix elements are now available; they were thoroughly 
cross-checked during the MC workshops held at FNAL and CERN during 2003. 
These programs generate samples of unweighted events in a standardized 
format (the LesHouches Accord \# 1) which can be passed to the parton 
shower-based MC programs to go from the partons to the real 
final-state hadrons. The matching between a LO multijet event generator 
and the parton shower MC suffers from the serious problems of double-counting 
and dependence on the parton-level cuts. For the case of $e^+ e^-$ collisions 
the problem has been solved beyond NLL accuracy with the CKKW algorithm. 
This can be extended to hadronic collisions, but the proof is still pending. 
However, recently, there has been an intense activity in its implementation 
on existing MC event generators going through many subtleties involved 
in the CKKW algorithm for hadronic collisions.

\vskip 12pt
The author wishes to thank the conveners of the Hard QCD Parallel Session 
for the invitation. M.L.~Mangano and A.D.~Polosa 
are gratefully aknowledged for many useful discussions and a careful reading of the 
manuscript.


\begin{thebibliography}{}
%
%
\bibitem{herwig}
G.~Marchesini and B.~R.~Webber,
Nucl.\ Phys.\ B {\bf 310} (1988) 461.
G.~Marchesini, et al., 
Comput.\ Phys.\ Commun.\  {\bf 67} (1992) 465.
G.~Corcella et al.,
JHEP {\bf 0101} (2001) 010
[hep-ph/0011363].
\bibitem{isajet}
F.~E.~Paige, S.~D.~Protopopescu, H.~Baer and X.~Tata,
hep-ph/9810440.
\bibitem{pythia}
T.~Sj\"ostrand, Comput.\ Phys.\ Commun.\  {\bf 82} (1994) 74; 
T.~Sj\"ostrand, et al., 
Comput.\ Phys.\ Commun.\  {\bf 135} (2001) 238.
\bibitem{acermc}
B.P.~Kersevan and E.~Richter-W\c{a}s, hep-ph/0201302.
\bibitem{alpgen}
M.L.~Mangano et al., JHEP \textbf{07} (2003) 001. \\ 
The documentation and the codes can be found at \\ 
{\tt http://mlm.home.cern.ch/mlm/alpgen/}.
\bibitem{amegic}
F.~Krauss, R.~Kuhn and G.~Soff, JHEP {\bf 0202} (2002) 044. 
\bibitem{comphep}
A.~Pukhov et al., hep-ph/9908288.
\bibitem{grace}
T.~Ishikawa et al. [MINAMI-TATEYA group Coll.], KEK-92-19; 
S.~Tsuno et al.,  Comput. Phys. Commun. {\bf 151} (2003) 216.
\bibitem{helac}
P.~D.~Draggiotis, R.~H.~Kleiss and C.~G.~Papadopoulos, Eur.\ Phy.\ J.\ C\ 
{\bf 24} (2002) 447; 
C.G.~Papadopoulos, talk given at \cite{mc4lhc}. 
\bibitem{madevent}
T.~Stelzer and W.F.~Long, Comput. Phys. Commun. \textbf{81} (1994) 357; 
F.~Maltoni and T.~Stelzer, JHEP {\bf 0302} (2003) 027. 
\bibitem{alpha}
F.~Caravaglios and M.~Moretti, 
Phys.\ Lett.\ B {\bf 358} (1995) 332; 
F.~Caravaglios, M.~L.~Mangano, M.~Moretti and R.~Pittau, 
Nucl.\ Phys.\ B {\bf 539} (1999) 215.
\bibitem{LesHouches}
E.~Boos et al., hep-ph/0109068.
\bibitem{gi}
M.~Veltman, Physica {\bf 29} (1963) 186; 
U.~Baur, J.A.M.~Vermaseren and D.~Zeppenfeld, Nucl.\ Phys.\ B {\bf 375} (1992) 3; 
Y.~Kurihara, D.~Perret-Gallix and Y.~Shimizu, Phy.\ Lett. B {\bf349} (1995) 367; 
U.~Baur and D.~Zeppenfeld, Phys.\ Rev.\ Lett.\ {\bf 75} (1995) 1002; 
R.G.~Stuart, Phys.\ Lett.\ B {\bf 262} (1991) 113; hep-ph/9603351; hep-ph/9706431; 
hep-ph/9706550; A.~Aeppli, G.J.~van~Oldenborgh and D.~Wyler, Nucl.\ Phys.\ B {\bf 428} 
(1994) 126; C.G.~Papadopoulos, Phys.\ Lett.\ B {\bf 352} (1995) 144; E.N.~Argyres et al., 
Phys.\ Lett.\ B {\bf 358} (1995) 339; W.~Beenakker et al., Nucl.\ Phys.\ B {\bf 500} 
(1997) 255; W.~Beenakker, F.A.~Berends and A.P.~Chapovsky, Nucl.\ Phys.\ B {\bf 548} 
(1999) 3; W.~Beenakker et al., hep-ph/0303105; G.~L\'opez Castro et al., Mod.\ Phys.\ 
Lett.\ A{\bf 40} (1991) 3679; A.~Denner et al., Nucl.\ Phys.\ B {\bf 560} (1999) 33; E.~Accomando, A.~Ballestrero and E.~Maina, Phys.\ Lett.\ B\ {\bf 479} 
(2000) 209. 
\bibitem{mlmtev} 
M.L.~Mangano, talk given at \cite{mctuning}, 15 November 2002.
\bibitem{ckkw}
S.~Catani et al., JHEP {\bf 0111} (2001) 063; L.~L\"onnblad, JHEP {\bf 0205} 
(2002) 046;  R.~Kuhn et al., hep-ph/0012025; F.~Krauss, R.~Kuhn and G.~Soff, 
J.\ Phys.\ G {\bf 26} (2000) L11.
\bibitem{durham}
S.~Catani, Y.L.~Dokshitzer and B.R.~Webber, Phys.\ Lett.\ B {\bf 285} (1992) 291; 
S.~Catani, Y.L.~Dokshitzer, M.H.~Seymour and B.R.~Webber, Nucl.\ Phys.\ B {\bf 406} 
(1993) 187.
\bibitem{krauss}
F.~Krauss, JHEP {\bf 0208} (2002) 015.
\bibitem{ckkwh} 
S.~Mrenna, talks given at \cite{mctuning,leshouchestev,mc4lhc};
P.~Richardson, talks given at \cite{mctuning,leshouchestev,mc4lhc}; 
F. Krauss, talk given at \cite{mctuning};  
A.~Sch\"alicke, talk given at \cite{mc4lhc}.
\bibitem{mlmmcw}
M.L.~Mangano, talks given at \cite{mctuning,mc4lhc}.
\bibitem{mctuning}
Monte Carlo Tuning Working Group, FNAL, 29-30 April 2003; 
{\tt http://cepa.fnal.gov/CPD/MCTuning/}.
\bibitem{leshouchestev}
Workshop ``Physics at TeV Colliders'', Les Houches, 26 May - 6 June 2003;\\ 
{\tt http://lappc-in39.in2p3.fr/conferences/LesHouches/}\\
{\tt Houches2003/}.
\bibitem{mc4lhc}
Workshop on ``MC tools for the LHC'', CERN, 7 July - 1 August 2003;\\ 
{\tt http://mlm.home.cern.ch/mlm/mcwshop03/mcwshop.html}.
\end{thebibliography}
%

\end{document}